\documentclass[11pt,twoside,a4paper]{article}

\makeatletter\if@twocolumn\PassOptionsToPackage{switch}{lineno}\else\fi\makeatother

\usepackage{amsfonts,amssymb,amsbsy,latexsym,amsmath,tabulary,graphicx,times,xcolor,subfig}

\usepackage[T1]{fontenc}

\usepackage{caption,fancyhdr,abstract,textcase}
\usepackage[paperheight=297mm,paperwidth=210mm,margin=30mm,left=30mm,right=30mm,top=30mm]{geometry}
\usepackage{textcase}
\usepackage{hyperref}
\usepackage{wrapfig, blindtext}
\usepackage[T1,T2A]{fontenc}
\makeatletter
\def\oupIndent{1pt}
\def\author#1{\gdef\@author{\hskip-\dimexpr(\tabcolsep)\hskip1pt\parbox{\dimexpr\textwidth-1pt}{\centering{\fontsize{13pt}{15.6pt}\selectfont  #1}}}}

\def\title#1{\gdef\@title{\vspace*{-3pc}\bfseries\centering\ifx\@articleType\@empty\else\@articleType\\\fi {\fontfamily{ppl}\fontsize{20pt}{24pt}\selectfont\MakeTextUppercase{ #1} \vspace*{24pt}}}}

\fancypagestyle{headings}{\fancyhf{}\fancyfoot[R]{}}

\fancypagestyle{plain}{\fancyhf{}\fancyfoot[R]{}}

\let\@articleType\@empty \def\articletype#1{\gdef\@articleType{{\normalfont\underline{#1}}}}

\def\abstractname{\textbf{\textit{A\small{BSTRACT}}}}

\renewenvironment{abstract} {\trivlist\item[]\leftskip\oupIndent\par\vskip4pt\noindent{\fontsize{13pt}{15.6pt}\selectfont\textit{\scshape\abstractname}}\mbox{\null}\vspace{5pt}\\ \itshape\fontsize{10pt}{12pt}\selectfont}{\noindent\endtrivlist}

\usepackage[explicit]{titlesec}
\def\NormalBaseline{\def\baselinestretch{1.1}}
\titleformat{\section}[hang]{\NormalBaseline\filright\large\scshape\bfseries\fontsize{14}{16.8}\selectfont}
{\fontsize{14}{16.8}\selectfont\thesection.}
{2pt}
{#1}
[]
\titleformat{\subsection}[hang]{\NormalBaseline\filright\bfseries\fontsize{12}{14.4}\selectfont}
{\thesubsection.}
{2pt}
{#1}
[]
\titleformat{\subsubsection}[hang]{\NormalBaseline\filright\bfseries\fontsize{11}{13.2}\selectfont}
{\thesubsubsection.}
{2pt}
{#1}
[]
\titleformat{\paragraph}[runin]{\NormalBaseline\filright\itshape\fontsize{11}{13.2}\selectfont}
{\theparagraph}
{2pt}
{#1}
[\unskip.]
\titleformat{\subparagraph}[runin]{\NormalBaseline\filright\fontsize{11}{13.2}\selectfont}
{\thesubparagraph}
{2pt}
{#1}
[\unskip.]

\titlespacing{\section}{0pt}{1.5\baselineskip}{.2\baselineskip}
\titlespacing{\subsection}{0pt}{1.5\baselineskip}{.2\baselineskip}
\titlespacing{\subsubsection}{0pt}{1.5\baselineskip}{.2\baselineskip}
\titlespacing{\paragraph}{0pt}{.5\baselineskip}{10pt}
\titlespacing{\subparagraph}{0pt}{.5\baselineskip}{10pt}

\makeatother

\date{}

\usepackage{url,multirow,morefloats,floatflt,cancel,tfrupee}
\makeatletter

\AtBeginDocument{\@ifpackageloaded{textcomp}{}{\usepackage{textcomp}}}
\makeatother
\usepackage{colortbl}
\usepackage{xcolor}
\usepackage{pifont}
\usepackage[nointegrals]{wasysym}
\makeatletter

\def\mcWidth#1{\csname TY@F#1\endcsname+\tabcolsep}

\def\cAlignHack{\rightskip\@flushglue\leftskip\@flushglue\parindent\z@\parfillskip\z@skip}
\def\rAlignHack{\rightskip\z@skip\leftskip\@flushglue \parindent\z@\parfillskip\z@skip}

\@ifundefined{etal}{}{}

\usepackage{ifxetex}
\ifxetex\else\if@twocolumn\@ifpackageloaded{stfloats}{}{\usepackage{dblfloatfix}}\fi\fi

\AtBeginDocument{
\expandafter\ifx\csname eqalign\endcsname\relax
\def\eqalign#1{\null\vcenter{\def\\{\cr}\openup\jot\m@th
  \ialign{\strut$\displaystyle{##}$\hfil&$\displaystyle{{}##}$\hfil
      \crcr#1\crcr}}\,}
\fi
}

\AtBeginDocument{%
  \@ifpackageloaded{endfloat}%
   {\renewcommand\efloat@iwrite[1]{\immediate\expandafter\protected@write\csname efloat@post#1\endcsname{}}}{\newif\ifefloat@tables}%
}%

\def\BreakURLText#1{\@tfor\brk@tempa:=#1\do{\brk@tempa\hskip0pt}}
\let\lt=<
\let\gt=>
\def\processVert{\ifmmode|\else\textbar\fi}

\@ifundefined{subparagraph}{
\def\subparagraph{\@startsection{paragraph}{5}{2\parindent}{0ex plus 0.1ex minus 0.1ex}%
{0ex}{\normalfont\small\itshape}}%
}{}

\newcommand\role[1]{\unskip}
\newcommand\aucollab[1]{\unskip}
  
\@ifundefined{tsGraphicsScaleX}{\gdef\tsGraphicsScaleX{1}}{}
\@ifundefined{tsGraphicsScaleY}{\gdef\tsGraphicsScaleY{.9}}{}
\def\checkGraphicsWidth{\ifdim\Gin@nat@width>\linewidth
	\tsGraphicsScaleX\linewidth\else\Gin@nat@width\fi}

\def\checkGraphicsHeight{\ifdim\Gin@nat@height>.9\textheight
	\tsGraphicsScaleY\textheight\else\Gin@nat@height\fi}

\def\fixFloatSize#1{}%\@ifundefined{processdelayedfloats}{\setbox0=\hbox{\includegraphics{#1}}\ifnum\wd0<\columnwidth\relax\renewenvironment{figure*}{\begin{figure}}{\end{figure}}\fi}{}}
\let\ts@includegraphics\includegraphics

\def\inlinegraphic[#1]#2{{\edef\@tempa{#1}\edef\baseline@shift{\ifx\@tempa\@empty0\else#1\fi}\edef\tempZ{\the\numexpr(\numexpr(\baseline@shift*\f@size/100))}\protect\raisebox{\tempZ pt}{\ts@includegraphics{#2}}}}

\AtBeginDocument{\def\includegraphics{\@ifnextchar[{\ts@includegraphics}{\ts@includegraphics[width=\checkGraphicsWidth,height=\checkGraphicsHeight,keepaspectratio]}}}

\DeclareMathAlphabet{\mathpzc}{OT1}{pzc}{m}{it}

\def\URL#1#2{\@ifundefined{href}{#2}{\href{#1}{#2}}}

\def\UrlOrds{\do\*\do\-\do\~\do\'\do\"\do\-}%
\g@addto@macro{\UrlBreaks}{\UrlOrds}

\edef\fntEncoding{\f@encoding}

\makeatother

\newif\ifmultipleabstract\multipleabstractfalse%
\usepackage{booktabs}
\usepackage[numbers,sort&compress]{natbib}

\fancypagestyle{plain}{%
	\fancyhead[C]{\small International Journal on Cryptography and Information Security (IJCIS), Vol. 10, No.3, Jul 2020}
	
}

\begin{document}

\title{GIVING UP PRIVACY FOR SECURITY: A SURVEY ON PRIVACY TRADE-OFF DURING PANDEMIC EMERGENCY}
\author{Sajedul Talukder\textsuperscript{1} and
        Md. Iftekharul Islam Sakib\textsuperscript{2} and
        Zahidur Talukder\textsuperscript{3} ~\\[12pt]\normalsize 
~\\ \fontsize{11}{13}\selectfont\textsuperscript{1}{Department of Mathematics and Computer Science\unskip, Edinboro University}\\\href{mailto:stalukder@edinboro.edu}{\tt stalukder@edinboro.edu}
~\\ \fontsize{11}{13}\selectfont\textsuperscript{2}{Department of Computer Science\unskip, University of Illinois at Urbana-Champaign}\\\href{mailto:misakib2@illinois.edu}{\tt misakib2@illinois.edu}
~\\ \fontsize{11}{13}\selectfont\textsuperscript{3}{Department of Computer Science\unskip, University of Texas at Arlington}\\\href{mailto:zahidurrahim.talukder@mavs.uta.edu}{\tt zahidurrahim.talukder@mavs.uta.edu}}

\def\journalTitle{International Journal on Cryptography and Information Security (IJCIS)}

\maketitle

\begin{abstract}
While the COVID-19 pandemic continues to be as complex as ever, the collection and exchange of data in the light of fighting coronavirus poses a major challenge for privacy systems around the globe. The disease's size and magnitude is not uncommon but it appears to be at the point of hysteria surrounding it. Consequently, in a very short time, extreme measures for dealing with the situation appear to have become the norm. Any such actions affect the privacy of individuals in particular. For some cases, there is intensive monitoring of the whole population while the medical data of those diagnosed with the virus is commonly circulated through institutions and nations. This may well be in the interest of saving the world from a deadly disease, but is it really appropriate and right? Although creative solutions have been implemented in many countries to address the issue, proponents of privacy are concerned that technologies will eventually erode privacy, while regulators and privacy supporters are worried about what kind of impact this could bring. While that tension has always been present, privacy has been thrown into sharp relief by the sheer urgency of containing an exponentially spreading virus. The essence of this dilemma indicates that establishing the right equilibrium will be the best solution. The jurisprudence concerning cases regarding the willingness of public officials to interfere with the constitutional right to privacy in the interests of national security or public health has repeatedly proven that a reasonable balance can be reached.
\end{abstract} \vspace*{12pt}\def\keywordstitle{Keywords}

\smallskip\noindent\textbf{\textit{\fontsize{13pt}{15.6pt}\selectfont K\small{EYWORDS}}}{\newline{\fontsize{10pt}{12pt}\selectfont \textit{Data Privacy, Privacy Trade-off, Pandemic Emergency.}}}
    
\section{\textbf{Introduction}} \label{sec:intro}

Today the coronavirus epidemic is overtaking areas of Asia, Europe and North America in full intensity, with the USA reporting more coronavirus-related deaths than any other country officially~\cite{fernandes2020economic}. Everyday life has been impacted by the proliferation of virus, from social networks~\cite{talukder2020tsc,ahani2020coronavirus,talukder2018abusniff,merchant2020social,talukder2019detection} to e-governance~\cite{sharfuddin2020.1760498,TSRICIEV14}, from digital automation~\cite{okereafor2020tackling,talukder2017usensewer} and cyber security~\cite{ahmad2020corona,talukdersurvey} to cellular networks~\cite{lajous2010mobile}. During the global coronavirus epidemic, many high-tech approaches are placed in motion to help mankind combat the virus. The entire world is coming to a standstill as policymakers are urging people to stay indoors, even using coercion. It would not have been possible to separate ourselves from civilization if people had no access to modern technology. Digital media services have provided content for hundreds of millions of users, a vast amount of individuals have begun telecommuting to schools and jobs, and the usage of safety devices has grown dramatically when millions more patients are diagnosed with telehealth applications - the last thing politicians across the globe expected was to see citizens with extremely infectious illness symptoms.

Nevertheless, the use of such technologies often poses several questions regarding privacy~\cite{talukder2020tools}. Israel also agreed to use the cellphone data gathered anonymously, typically meant for counter-terrorism use. The information was used to identify people who crossed paths with persons bearing COVID-19. A combination of geotracking and AI technology has allowed the Israeli government to recognize individuals who should be quarantined because of their potential coronavirus exposure. When the coronavirus pandemic spread across the globe, countries leveraged massive monitoring networks to track the transmission of the virus and pressured policymakers around the world to consider the trade-offs with millions of people in public health and safety. Government agencies in South Korea are harnessing surveillance camera video, mobile location data, and credit card payment records to help monitor recent coronavirus patient movements and create virus transmission chains. Iran launched an app promising to cure COVID-19. Today, much of what it does is collect millions of people's location info, effectively serving as a way for the government to monitor its population in real time.

The U.S. aims to counter coronavirus through mobile location info. The White House and the Centers for Disease Control and Prevention are calling on Facebook, Google and other tech giants to give them greater access to American smartphone location data to help them fight the spread of coronavirus, according to four people at the companies involved in the discussions who are not authorized to speak publicly about it. Advocates for privacy are concerned. Although, some sources emphasized that the effort would be anonymized and the government would not have access to the locations of specific individuals. Federal health authorities say they will use confidential, aggregated consumer data obtained by the internet firms to monitor the distribution of the virus - a technique known as ``syndromic surveillance'', and avoid more outbreaks. They might still use the data to see if people were using ``social distancing.'' In China, South Korea and Israel, similar and more extreme monitoring techniques have also been used. The moves set off warning bells for privacy activists who are fearful of what the government might do with consumer data. Facebook also offers anonymized data to health experts and Charities in certain countries to help deter epidemic efforts. Yet other reports cautioned that supplying government with greater access to anonymized location data might now contribute to the downline degradation of individual privacy, especially if the government begins asking for non-anonymized data.

The data-protection consequences of efforts to monitor coronavirus dissemination have been increasingly evident in Asia. When this economic epidemic is fully under control, it is possible that China will be commended for the scientific success with which it ultimately succeeded in preventing an outbreak that could have infected billions. Yet it sure does come at a expense. There are examples of ``epidemic maps'' displaying the exact position in real time of reported and alleged cases so people can stop traveling to the same locations. There's also an app that lets people communicate with someone who acquired the virus whether they've been on a train or plane. Such interventions are also successful, but they involve the vast processing and distribution of accurate medical data. Similarly, creative approaches to addressing the problem have been introduced in Korea and Singapore, and ultimately, given the broad and apparent intrusions to privacy, they seem to have been successful.

Nowhere in the world, the right to privacy is an absolute right. Laws on privacy and data security can not and do not mess with a common-sense solution to saving lives. Of this reason, all these systems allow the use and sharing of data for that purpose where appropriate. Around the same time, it is not possible to ignore the criteria laid down in the legislation-except in times of crisis. Disproportionate decisions and measures are often the result of knee-jerk reactions, and everybody is at risk when that happens on a global scale-no matter how often you wash your hands.
\\ 

\noindent
{\bf Our Contributions}:
This paper presents the following contributions:

\begin{itemize}
	
\item
{\bf Survey on privacy issues during pandemic}.
This paper provides the first comprehensive survey on various privacy issues on various fronts that are taking place during the pandemic. To the best of our knowledge, this is the first comprehensive survey that addresses privacy issues from all sectors from surveillance to medical data.
	
\item
{\bf Privacy principles recommendation}.
This paper presents various privacy principles recommendation for monitoring app and device designers that include privacy-conscious mechanisms, aggregated anonymized data and volunteered data.
	
\item
{\bf Guideline for post-pandemic privacy restoration}.
Finally, this paper presents a comprehensive guideline to ensure post-pandemic privacy restoration that would be beneficial for the public and private sectors alike.

\end{itemize}

The rest of the paper is organized as follows. Section II describes the privacy issues. Section III describes the recommended privacy principles. Section IV discusses the post-pandemic privacy restoration. Finally, Section V concludes the paper with a highlight on the scope of future work.

\section{\textbf{Privacy Issues}}

\subsection{Aggressive Surveillance Measures}
Deutsche Telekom, Europe's biggest telco firm, announced that it is turning over 5 GB of consumer data to the German Robert Koch Institute to counter coronavirus transmission, COVID-19. The Institute would use the anonymized data to monitor the activities of the general public to forecast how the virus spreads and to help address questions about the social distance effectiveness. Similarly, Vodafone has published a five-point program in which it promises that private consumer data will now be transferred to Lombardy, Italy. And Austria's largest telco, A1, has contributed the data as well. Critics point out that other countries are already making more authoritarian use of mobile-phone data. These data are used in China, Israel, and South Korea to monitor the connections of contaminated locals and to ensure quarantine is enforced. Critics also question the legality of the donation, and whether the data privacy of customers has been respected - and whether the donation of data would be useful. Although GPS-related data, such as that obtained by Google, can be very reliable, cellular providers' mobile phone tracking data still use cell phone towers to locate the customers~\cite{talukder2017attacks}. Its accuracy ranges from 25 to 100 metres, that might not be particularly helpful in big cities. While European officials assured data-privacy advocates of not tracking individuals, consent and transparency issues were not eased. Much information is associated with the mobility data, which could be worse with the lack of transparency of the Telecom companies that have long been collecting and selling geolocation data of their customers~\cite{TC2017,talukder2019ccsc}. In the US, the government is currently engaged in talks with tech giants such as Google, Apple, and Facebook about how data from their customers could be used to prevent COVID-19 from spreading.

\begin{figure}[t]
\centering
\includegraphics[width=.8\columnwidth]{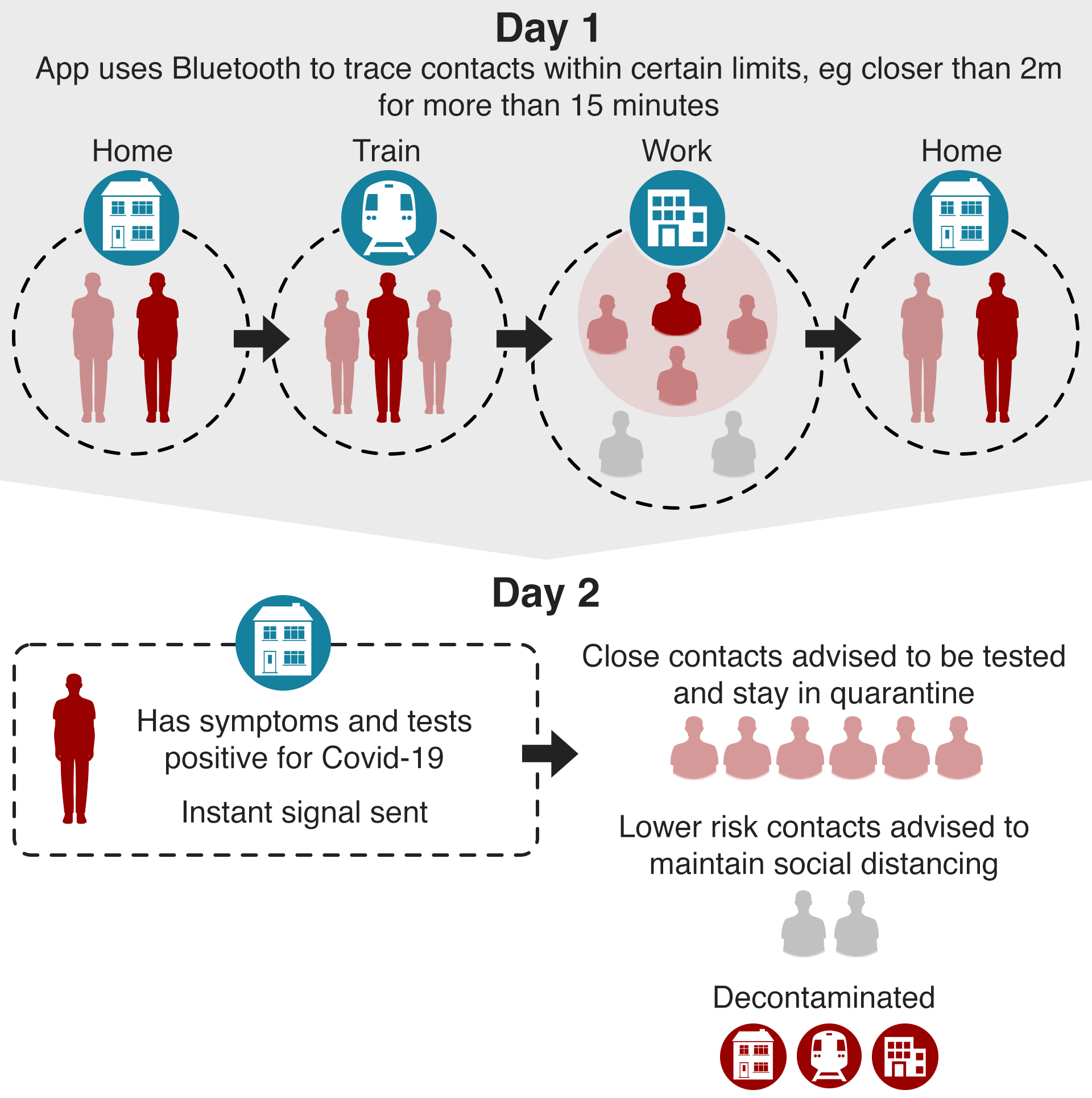}
\caption{Oxford University contact-tracing app.}
%\vspace{-15pt}
\label{fig:app}
\end{figure}

A medical school in Hanover in Germany is collaborating with local mapping firm, Ubilabs, to develop an app that will enable individualized monitoring of infections. A person who has tested positive for COVID-19 voluntarily donates the GPS data~\cite{TSRICEEICT14} from their phone with this app, named GeoHealth and expected to be available in a few weeks time. Many users would be able to say whether they are at the same location, at the same time, as the infected person. If the users get a ``red signal,'' they're told to go and get checked, warning them they were really close by. Recently, the government of the autonomous Spanish territory of Catalonia has released its own, related app, named STOP COVID19 CAT. Anonymous data can also be quickly re-identified, which, they claim, raises fear in cases like this. This, and the absence of a system, has in the past been a significant obstacle to the use of evidence in a humanitarian situation, as happened during Ebola. The Oxford University academics and the health service's digital innovation unit in UK also designed a contact-tracing app (see Fig~\ref{fig:app}) that takes into account different age groups, household structures and movement patterns in an effort to try to maximise the number of people who could be allowed to freely move~\cite{BBCapp}.

When South Korea is fighting a snowballing number of COVID-19 incidents, the government is letting people know if they are in a patient's area. But the volume of information has led to some awkward moments and there is just as much fear of social stigma as of illness. Such warnings arrive all day, all day, asking people where an infected person was. People may also visit the Ministry of Health and Welfare website for information. When online searching for the case number of a virus victim, the associated inquiries include ``contact information,'' ``name,'' ``photo,'' ``father'' or even ``adultery.'' There are no names or addresses given but certain people nevertheless tend to connect the dots and recognise individuals. And, even though patients aren't specifically known, online they risk criticism or mockery. Doctors warn that pursuing patients online may have very serious consequences. Malicious online remarks have long been a issue in South Korea, and have led to suicide in some situations. Another known thing is elevated levels of anxiety due to online criticism and endured sleep deprivation. Because the virus spreads quickly, the knowledge they need to defend themselves and others will be made available to the public.

The good news is, however, that data security agencies from around the world are coming in to provide their advice and feedback on data management practices and on fighting coronavirus. The contrasting perspective among the opinions of data protection agencies - who may be categorized as punitive, impartial, or acceptable - indicates that the best response would be to pursue a reasonable middle ground that does not neglect the enforcement of basic standards of privacy. This is also in accordance with the announcement released on March 16 by the European Data Protection Board (EDPB) stressing that data security laws (such as GDPR) do not obstruct steps taken to counter the coronavirus pandemic. Table~\ref{tab:data} lists the guidelines on data management practices and on fighting coronavirus issued by various countries and organizations.

\begin{table*}[!htbp]
  \centering
  \resizebox{\linewidth}{!}{
    \begin{tabular}{|l|p{33em}|l|p{33em}|}
    \toprule
    \rowcolor{gray!20} {\textbf{Country}} & \textbf{Data protection and Coronavirus (COVID-19) resources} & {\textbf{Country}} & \textbf{Data protection and Coronavirus (COVID-19) resources}  \\
    \midrule
    \multicolumn{1}{|c}{{\textbf{Albania}}} & IDP  & \multicolumn{1}{c}{{\textbf{Luxembourg}}} & \textbf{National Commission for Data Protection} \\
          & Guidelines on the protection of personal data in the context of the measures taken against COVID-19 [EN] &       & Coronavirus (COVID-19): recommendations by the CNPD on the processing of personal data in the context of a health crisis [EN] [FR] \\
    \midrule
    \multicolumn{1}{|c}{{\textbf{Argentina}}} & Agencia de Acceso a la Información Pública & \multicolumn{1}{c}{{\textbf{Mexico}}} & \textbf{National Institute for Transparency, Access to Information and Personal Data Protection (INAI)} \\
          & Tratamiento de datos personales ante el Coronavirus [ES] &       & Statement - Ante casos de COVID-19, INAI emite recomendaciones para tratamiento de datos personales [ES] \\
\midrule   \multicolumn{1}{|c}{{\textbf{Australia}}} & Office of the Australian Information Commissioner (OAIC) &       & Statement - Suspende INAI eventos públicos, por recomendación de la SSA para evitar contagio de COVID-19 [ES] \\
          & Coronavirus (COVID-19): Understanding your privacy obligations to your staff - Agencies [EN] &       & Statement - Adoptará INAI como medida de prevención el trabajo a distancia ante COVID-19 [ES] \\
\midrule    \multicolumn{1}{|c}{{\textbf{Austria}}} & Austrian Data Protection Authority &       & \textbf{Transparency, Public Information Access and Personal Data Protection Institute of Estado de México and Municipalities (Infoem)} \\
          & Data Protection Authority Information on Coronavirus (COVID-19) [DE] &       & Recomendaciones respecto a la garantía de los derechos de protección de datos personales y acceso a la información pública, ante brote de COVID-19 [ES] \\
    \midrule
    \multicolumn{1}{|c}{{\textbf{Bulgaria}}} & Commission for Personal Data Protection & \multicolumn{1}{c}{{\textbf{Netherlands}}} & Data Protection Commission (Autoriteit Persoonsgegevens) \\
          & КЗЛД въвежда противоепидемични мерки срещу разпространението на COVID-19 [BG] &       & Statement - AP geeft organisaties meer tijd vanwege coronacrisis [NL] \\
\midrule   \multicolumn{1}{|c}{{\textbf{Canada}}} & \textbf{Office of the Privacy Commissioner of Canada} &       & FAQs on testing employees for Coronavirus (Corona op de werkvloer) [NL] \\
          & Announcement: Commissioner issues guidance on privacy and the COVID-19 outbreak [EN] &       & FAQs on working from home (Veilig thuiswerken tijdens corona) [NL] \\
\midrule         & Guidance: Privacy and the COVID-19 outbreak [EN] & \multicolumn{1}{c}{{\textbf{New Zealand}}} & Office of the Privacy Commissioner \\
          & \multicolumn{1}{l|}{} &       & COVID-19 and privacy FAQs [EN] \\
\midrule          & \textbf{Office of the Information and Privacy Commissioner of Alberta} & \multicolumn{1}{c}{{\textbf{Perú}}} & \textbf{Autoridad Nacional de Protección de Datos Personales del Perú} \\
          & Privacy in a Pandemic [EN] &       & Divulgar datos personales de pacientes con coronavirus puede ser multado hasta con 215 mil soles [ES] \\
    \midrule
    \multicolumn{1}{|c}{{\textbf{Czech Republic}}} & Office for Personal Data Protection & \multicolumn{1}{c}{{\textbf{Philippines}}} & \textbf{National Privacy Commission} \\
          & Ke zpracování osobních údajů v rámci opatření proti šíření koronaviru [CS] &       & Data protection in times of Emergency [EN] \\
\midrule
    \multicolumn{1}{|c}{\textbf{EDPB}} & Statement of the EDPB Chair on the processing of personal data in the context of the COVID-19 outbreak [EN] &       & Protecting personal data in the time of COVID-19 [EN] \\
    \midrule
    \multicolumn{1}{|c}{{\textbf{Finland}}} & Office of the Data Protection Ombudsman & \multicolumn{1}{c}{{\textbf{Poland}}} & \textbf{Personal Data Protection Office of Poland} \\
          & Data protection and limiting the spread of coronavirus [EN] &       & Statement by the President of the Personal Data Protection Office on coronavirus [EN] \\
    \midrule
    \multicolumn{1}{|c}{{\textbf{France}}} & Commission Nationale de l’Informatique et des Libertés & \multicolumn{1}{c}{{\textbf{San Marino}}} & Autorità Garante per la protezione dei dati personali \\
          & Coronavirus (COVID-19) : les rappels de la CNIL sur la collecte de données personnelles [FR] &       & Public announcement on COVID-19 emergency [EN] [IT] \\
    \midrule
    \multicolumn{1}{|c}{{\textbf{Gibraltar}}} & Gibraltar Regulatory Authority & \multicolumn{1}{c}{{\textbf{Slovakia}}} & Office for Personal Data Protection of the Slovak Republic \\
          & Data protection and Coronavirus: What you need to know [EN] &       & Statement of the EDPB Chair on the processing of personal data in the context of the COVID-19 outbreak [SK] \\
\midrule   \multicolumn{1}{|c}{{\textbf{Germany}}} & Office of the Federal Commissioner for Data Protection and Freedom of Information &       & Coronavirus and processing of personal data [SK] \\
\midrule          & DSK provides information on data protection and Coronavirus [DE] & \multicolumn{1}{c}{{\textbf{Spain}}} & \textbf{Catalan Data Protection Authority} \\
          & German Data Protection Supervisory Authorities joint information paper on data protection and the Coronavirus pandemic [DE] &       & Statement regarding the processing of personal data related to the measures to deal with COVID-19 [CA][EN] \\
          & \multicolumn{1}{l|}{} &       & \multicolumn{1}{l|}{} \\
          & \textbf{Mecklenburg-West Pomerania: Data Protection Commissioner (Landesbeauftragte für den Datenschutz Mecklenburg-Vorpommern)} &       & \textbf{Spanish Data Protection Agency (Agencia Española de Protección de Datos)} \\
          & Statement on Privacy and Coronavirus [DE] &       & Report from the State Legal Service Department (the Spanish DPA) on Processing Activities Relating to the Obligation for Controllers from Private Companies and Public Administrations to Report on Workers Suffering from COVID-19 [EN][ES] \\
\midrule    \multicolumn{1}{|c}{{\textbf{Hong Kong}}} & Privacy Commissioner for Personal Data &       & COVID-19 FAQs [ES] \\
          & The Use of Information on Social Media for Tracking Potential Carriers of COVID-19 [EN] &       & Statement - La AEPD publica un informe sobre los tratamientos de datos en relación con el COVID-19 [ES] \\
          & Masks and Police Officers’ Car Registration Numbers [EN] &       & Statement - Comunicado de la AEPD en relación con webs y apps que ofrecen autoevaluaciones y consejos sobre el Coronavirus[ES] \\
          & Privacy Commissioner Responds to Privacy Issues Arising from Mandatory Quarantine Measures and Provides Updates on Doxxing [EN] &       & Statement - Campañas de phishing sobre el COVID-19 [ES] \\
    \midrule
    \multicolumn{1}{|c}{{\textbf{Hungary}}} & Hungarian National Authority for Data Protection and Freedom of Information & \multicolumn{1}{c}{{\textbf{Switzerland}}} & \textbf{Federal Data Protection and Information Commissioner (FDPIC)} \\
          & Information on processing data related to the Coronavirus epidemic [EN] &       & Data protection legal framework for the containment of the coronavirus [EN][DE][FR][IT] \\
    \midrule
    \multicolumn{1}{|c}{{\textbf{Ireland}}} & Data Protection Commission & \multicolumn{1}{c}{{\textbf{United Kingdom}}} & \textbf{Information Commissioner's Office (ICO)} \\
          & Data Protection and COVID-19 [EN] &       & Data protection and coronavirus: statement for health and care practitioners [EN] \\
\midrule    \multicolumn{1}{|c}{{\textbf{Italy}}} & \textbf{Data Protection Commission (Garante per la protezione dei dati personali)} &       & COVID-19: general data protection advice for data controllers [EN] \\
\midrule         & Coronavirus and data protection [IT] & \multicolumn{1}{c}{\textbf{United Nations}} & COVID-19: States should not abuse emergency measures to suppress human rights - UN experts [EN] \\
    \midrule
    \multicolumn{1}{|c}{{\textbf{Jersey}}} & \textbf{Office of the Information Commissioner} & \multicolumn{1}{c}{{\textbf{United States}}} & \textbf{Federal Trade Commission (FTC)} \\
          & Data Protection and Coronavirus [EN] &       & FTC, FDA Send Warning Letters to Seven Companies about Unsupported Claims that Products Can Treat or Prevent Coronavirus [EN] \\
\midrule   \multicolumn{1}{|c}{{\textbf{Lithuania}}} & State Data Protection Inspectorate &       & FERPA \& Coronavirus Disease 2019 (COVID-19) \\
          & Personal Data Protection and Coronavirus COVID-19 [EN] [LT] &       & Coronavirus Scams: What the FTC is Doing [EN] \\
    \bottomrule
    \end{tabular}%
    }
    \vspace{5pt}
  \caption{Country level data protection and management guidelines.}
  \label{tab:data}
\end{table*}

\subsection{Privacy of Medical Data}
The coronavirus epidemic is forcing the US government to relax one of its few laws relating to data protection. The Health Insurance Portability and Accountability Act is one of the protections against the misuse of patient records, but the Health and Human Services Administration said it would waive fines for suspected HIPAA breaches which is very common~\cite{talukder2018mobile}. HIPAA prevents patients from sharing their personal records with health care providers, thereby restricting the exploitation of the records. Such privacy provisions restrict what types of technologies health care facilities can use, but this is evolving with the coronavirus pandemic. The Department of Health and Human Services (HHS) has released a Notice to ensure that the organizations affected by HIPAA and their employees are aware of this data rights provision that shall not be set aside in the case of an emergency. Although HIPAA safeguards the protection of ``protected health information'' (PHI), this is adjusted to ensure that reasonable uses and releases of the information can also be made as required to handle a individual (the employee or dependent), to protect the general welfare of the country, and to avoid a significant and immediate danger to a person or the public's health and safety. Via remote communication systems, protected health care providers subject to the HIPAA Regulations may try to connect with patients and offer telehealth services. Almost all video messaging provider complies with the HIPAA, as are specialized systems such as Zoom for Healthcare and Skype for Business. However, with the threat of COVID-19, the coronavirus-induced respiratory disease, to spread at a rapid pace, and with governments advising people to remain home to control the epidemic, HHS has agreed to open more popular video chatting apps for doctors to use. This includes famous FaceTime, Facebook Messenger, Google Hangouts, and Skype apps. The Civil Rights Office of the HHS said that with such non-compliant video messaging systems, it would not place restrictions on health care providers.

Another concerning thing from the Bulletin notes that, in an outbreak of an infectious illness such as COVID-19, HIPAA-covered workers will have the same right as HIPAA-excluded workers to share workplace information with others if necessary to prevent or minimize a significant and urgent risk to the health and safety of a worker or the public in keeping with the applicable regulations. Therefore, employers may share information on the location, general illness, or death of an employee or child to identify, locate, and warn family members, guardians, and other individuals responsible for the care of that individual, if necessary. HIPAA forbids, without their permission, the release of information about the employee or dependent's illness to the public. But the issue is that the media and the general public are not bound by HIPAA regulations and are also not subject to HIPAA requirements because they provide details about a person who has contracted COVID-19. PHI typically does not provide personally identified health documents kept by the employer in employee reports necessary by the employer to meet its responsibilities under the FMLA, ADA and related legislation, as well as files or records pertaining to other work problems such as unemployment benefits and sick leave demands. So while there may be other applicable state or federal privacy rules, HIPAA generally does not apply if the information is not obtained from the health plan group. PHI may be needed by health authorities and others responsible for ensuring public health and safety to allow them to accomplish their mission of protecting the public from disease. The HIPAA privacy law also includes provisions which will require employers to exchange information about workers or dependents who have contracted COVID-19 with state and federal public health agencies, such as the Centers for Disease Control and Prevention (CDC) and state and local health departments.

\subsection{Privacy of Student Data}
FERPA is a Federal law which protects the privacy of records of student education. (20 U.S.C. § 1232 g; 34 C.F.R. Section 99) All government agencies and organizations seeking funds under any system operated by the Secretary of Education shall be subject to law. Within FERPA, the word ``educational agencies and institutions'' typically encompasses in the primary level school districts and high schools. In slowing the proliferation of COVID-19 in U.S. populations, educational organizations and institutions such as school boards, hospitals, colleges, and universities will play a major role. Educational organizations and institutions may better protect their schools and neighborhoods by knowledge exchange and collaboration with public health departments. The term ``PII'' refers to the name or identification number of a student, as well as other information which can be used to distinguish or trace the identity of an individual, either directly or indirectly, through links with other information. FERPA forbids government authorities (e.g., school districts) and organizations ( e.g., schools) from releasing PII from the academic record of the children without prior notification except where an exception to the general consent provision applies to FERPA. For example, under one such exemption, ``health or safety emergency,'' educational agencies and universities may report PII from student education records to a public health agency, without prior written permission in connection with an emergency, if disclosure of the details by the public health authority is required to protect the health or safety of students or other individuals. FERPA requires school agencies and organizations to reveal information from educational documents without permission until all PIIs have been deleted, provided that the organization or institution has made a fair decision that the student's name is not publicly identifiable, whether by single or multiple releases, and taking into account all fairly available information. Therefore, reporting that a student in a certain class or grade level is missing would be troublesome because, for example, there is a directory containing the names of every student in that class or grade. Many known attacks have occurred and can de-anonymize individual data from the anonymized data collection.

There is an exception to the general consent requirement of FERPA, ``health or safety emergency,'' which is limited in time to the emergency period and generally does not allow for a blanket release of PII from student education records. However, the concern is that while educational agencies and institutions may frequently address threats to the health or safety of students or other persons in a manner that does not identify a particular student, FERPA allows educational agencies and institutions to disclose PII from student education records to relevant parties in connection with an emergency without prior written consent. Usually, law enforcement officers, public health authorities, qualified medical professionals, and parents (including parents of an qualifying student) are the categories of interested parties to which PII may be reported from school documents under this FERPA exemption. The decision by an educational organization or entity that there is a real emergency is not dependent on a vague or remote warning of a potential or imminent disaster for which the possibility of occurrence is unclear, as should be discussed in general emergency preparedness procedures, for the purposes of FERPA's health or safety emergency exemption.If local public health agencies decide that a public health emergency, such as COVID-19, is a serious danger to the city's children or other persons, an educational organization or entity within the region may conclude that an emergency also occurs. Under the FERPA health or safety emergency provision, it is the duty of an educational organization or institution to make a case-by-case decision as to whether to report PII from educational documents, and it must take into account the entirety of the circumstances of the hazard. When the school organization or entity decides that there is an articulate and important danger to the student's or other individual's health or safety, and whether other parties require the PII from educational documents to protect the student's or other individual's health or safety, it may reveal the information without permission to those parties. It is a robust principle by which the Department does not substitute its opinion with that of an educational agency or institution such that the educational agency or institution will have sufficient services to sustain the situation because, on the basis of the knowledge available at the time of the assessment of the educational agency or institution, there is a reasonable reason for such a decision. Therefore, there is a substantial possibility that certain school organizations and institutions may report to the Department of Public Health the names, addresses and telephone numbers of missing students without permission, so that the Department of Health can contact their parents to determine the illnesses of the children. There is also a possibility that a single pupil, teacher, or other school official may be reported by the school as requiring COVID-19 for parents of other students in the organization.

\subsection{Teleconferencing Privacy (Zoombombing)}
As a result of its coronavirus-fuelled market expansion, Teleconferencing firm Zoom is attracting attention. Internet-rights activists claim they want Zoom, which was flooded by demand after the ebola epidemic, to begin releasing a transparency report outlining its data protection policies, and how it responds to governments trying to clamp down on freedom of speech. The growing need for their services makes Zoom a priority for third parties, ranging from law enforcement to sophisticated hackers targeting confidential data and sensitive data. Meanwhile such gatherings will attract attention from regulators trying to regulate the flow of information as people assemble online.

While the video conference software Zoom is gaining traction as a result of increased usage of the coronavirus pandemic, federal authorities are now warning of a new possible risk for privacy and security dubbed ``Zoombombing.'' The phrase applies to a type of cyber abuse reported by some users of the app, who have claimed that some of their calls have been intercepted by anonymous persons and trolls speculating hatefully. ``Zoombombing'' has become so widespread that the FBI has released a news release reminding people of the threat which includes profanity, pornography and swastika. These ``Zoombombing'' incidents occur as Zoom faces criticism for its privacy rights, which customers, technology experts and US regulators have flagged. In their efforts to mitigate these threats from potential hackers or trolls, federal officials urged those using video teleconferencing apps to exercise due diligence and caution. Malicious cyber actors are searching for ways to manipulate flaws in telework applications to obtain private information, eavesdrop on phone calls or simulated meetings, or conduct other disruptive activities.

\subsection{Privacy of Sensitive Data}
Coronavirus is a particularly infectious illness and will most likely entail the monitoring of billions of individuals during 2020. Patients also exchange DNA-like evidence by handing out a swab sample to hospitals. Although there's no doubt that governments will do the best for their people, it's still important to remember that in the past, there have been incidents of government records compromised. DOJ filed charges against Chinese hackers for taping into Equifax just a few months earlier. Knowing that most Americans would not be as pleased if important DNA data gathered through COVID-19 research ends up in the hands of foreign states such as China and Iran, is necessary.

Clearview AI, a facial recognition company claiming to have scraped billions of viral images off the internet and built tools capable of recognizing a face in seconds. It sells itself to law enforcement within the U.S. but also to designated oppressive governments around the world as part of a global growth program with histories of human rights violations. Palantir, however, has comprehensive law enforcement contracts and offers little or no accountability about its activities unless you are a client. The development of public-private partnerships to exchange confidential data with organizations such as Clearview AI and Palantir in times of disaster, such as a terrorist incident or pandemic, provides short-term gains but has an disturbing effect on data protection even after the emergency passes. One major concern is that after the crisis passes, new surveillance technologies deployed during the coronavirus crises will become the'new normal' and permanently embedded in daily life. It will result in ongoing mass population monitoring without proper oversight, accountability or justice. Of this, there is a history, albeit not long after. The terrorist attacks of 9/11 in 2001 led to the proliferation of surveillance cameras and networks across the U.S. and the Patriot Act, a new statute that eliminated statutory guardrails for government oversight and limited accountability, increasing the invasive and extensive surveillance powers of the National Security Agency later exposed by whistleblower Edward Snowden. Despite the public uproar against NSA activities, it is yet to be de-authorized by lawmakers.

There is no question about the presence of government departments to represent the people. Nonetheless, people run institutions and often people make errors, errors that can reveal sensitive information to hackers who do not have access to it. Governments will certainly ensure that the tools they have built to combat extremism, such as the position monitoring device used in Israel, remain in safe hands only. All people who want to secure their digital life need to ensure that all connected devices are protected by reliable anti-virus software as regards regular folks.	
	
\subsection{Privacy of Tech Company Data}
After the advent of the coronavirus pandemic, tech companies have provided their resources, funds, and face mask stashes to help. Many organizations that work with the data are now stepping up providing their tools for data collection to try and monitor or avoid the virus spread. Unacast, a data company that collects and provides the retail, real estate, marketing, and tourism industries with cellphone location data and analysis, has recently revealed something called the Social Distancing Scoreboard~\cite{Unacast}. The scoreboard is an interactive map that gives letter ratings to each state and county in America depending on how often Unacast's examination of data conclude that its people conduct social distancing. It's the first offering from the company's latest COVID-19 Location Data Toolkit, and more location data will be introduced in the coming days and weeks to demonstrate trends and patterns that the company hopes would.

Unacast isn't the only tech firm that has used the technology for what it claims is a public benefit these days. The ``Data for Good'' platform on Facebook uses de-identified statistical data from its users to support its Disease Prevention Maps, which will offer information about where people live and where they travel and can help health agencies monitor disease transmission or anticipate when they will be heading next. Kinsa Health uses data from its smart thermometers to attempt to identify extremely high rates of influenza with its US Health Climate Monitor, which the organization claims has forecast the spread of flu correctly in the past and may also map outbreaks of coronavirus. Unacast is a bit different though. For the Facebook system and the Kinsa app, users will opt to monitor their position and instead have a clear interaction with those businesses. In comparison, Unacast receives app data from a number of third-party providers. Such outlets, according to its privacy policy, include Unacast's suppliers as well as the software development kit, or SDK, which it puts in applications.

The stumbling point is that users give one of those applications permission to access their location data without realizing that this location data will go to Unacast as well. To the average consumer, there is no simple way to see what SDKs an app may use, because the privacy rules of the app typically state the details goes to third parties without disclosing who certain parties are. Unacast claims its SDK is its ``preferred'' data base on its website, but when we asked for clarification the company wouldn't specify the applications or partners it deals with. An study by Apptopia, a smartphone app technology company, finds Unacast's SDK in all forms of iOS and Android phones, including smart TV remotes, time trackers, sports, wireless wifi locators, weather forecasters and phase trackers. For these apps, users can always turn the location tracking off but some of them obviously need the location services to be able to work at all.

While the chart itself can be a helpful device, it also makes such data collection activities even more visible behind the scenes — and how precise the data collected can become. Unacast is hardly the only company doing this kind of data crunching. For example, oneAudience, a marketing company, puts its SDK into apps to gather user information. The firm also illegally collected social media data, as Facebook alleged in a recent complaint, but the firm admitted this compilation was accidental, and that it changed its SDK to stop it. The dilemma is there's still no way for an ordinary customer swept up in the vortex of tech companies to know exactly what's going on with their position data, and which companies have access to it and which firms secure it properly. Broadly speaking, there are no federal regulations prohibiting the processing of such data and it is impossible for users to take advantage of the privacy protections they have because most of them do not even know that data collection services such as Unacast exist. Privacy activists have fears that the extreme nature of the pandemic could also cause privacy rights to erode in this region. The problem then is, in this case: Will the immediate gains of this data outweigh the risks of long term privacy? One can guess the obvious answer would be a ``No''.

\section{\textbf{Recommended Privacy Principles}}
One of the best tools at the hands of public health authorities during a epidemic crisis is low-tech detective work. If a person is infected with a disease such as COVID-19, the disease caused by the novel coronavirus, public health authorities figure out where they have been lately and monitor all people they have been in touch with. The tension between preserving the rights of people and obtaining knowledge that is important to the general interest varies over the course of dissemination of a epidemic. Core specifics would need to be sorted out by monitoring device designers: how to assess phone proximity and user safety status, where the information is processed, who uses it, and in what format. Digital contact tracing services are currently working in a variety of countries, but specifics are sparse, and questions over privacy exist. In the following, we propose several recommended privacy principles.

\subsection{\textbf{Privacy-conscious mechanisms}}
Several new initiatives are aimed at developing cooperative, privacy-conscious mechanisms for monitoring phones. A Massachusetts Institute of Technology team launched a prototype of a Private Package app: Safe Paths. The app stores up to 28 days of GPS location data provided by a user, logged every 5 minutes. If the individual performs successful coronavirus tests, they can want to share their recent data with health authorities in order to recognize and publicize the areas where others may have been at risk. A new version of the app, which will be announced early, will equate recent movements of a user with an infected person's route and warn them to possible contact. Users wouldn't know much more about the infected person — not their age, race, or geographical trajectory. The project, which includes colleagues from Harvard University and the Mayo Clinic, is debating the application's ongoing pilot trials with a dozen cities and nations in every corner of the country. Creating optional and anonymizing data contributions are healthy choices for protecting human rights. Though it's a safe way to do so lawfully, these devices can only minimize disease risk if they're used by enough people.

\subsection{\textbf{Aggregated anonymized data}}
For countries where data protection rules are stringent, one choice for data collection is to allow telephone and other software providers to exchange confidential, aggregated information that they have already gathered. The U.S. and European Union regulations are very clear on how consumers of software and smartphones have to agree to the processing of their data — and how much information businesses have to provide on how such data will be accessed, processed, and exchanged. Operating under those limits, mobile providers in Germany and Italy have started to exchange cell phone tracking data in an aggregated, anonymized format with health authorities. In Germany, which has some of the most robust data privacy safeguards in Europe, the government may force a technology provider to exchange location data on an person for national security purposes. While individual users are not known, the data can show general patterns about where and when people meet and prevent infection spreading. Google and Facebook are also debating the exchange of anonymised location details with the U.S. government.

\subsection{\textbf{Using volunteered data}}
Another choice is to use a coronavirus-specific app to start new, requiring users to willingly share their location and health details. One emerging app in Germany depends in part on location data already held by Google for its account holders. A person who is performing favorably may use the GeoHealth app to donate the history of their place. The data will then be anonymized and processed on a central server. A data analytics platform developed by the Ubilabs tech firm will equate the history of users' activity with that of infected individuals, and the system will display color-coded warnings depending on how recent they might have experienced the virus. As a mixture of GPS mapping, wireless network data, and telephone communications using Bluetooth, the device will be able to identify when a mobile comes within one meter from another.

\section{\textbf{Post-pandemic Privacy Restoration}}
Coordinated data-sharing has become an essential tool in the ongoing fight against coronavirus as scientists around the world work tirelessly to develop a viable vaccine. Mass data collection methods are already being put to use in an effort to establish effective public health strategies and protocols to curtail the spread of COVID-19. Naturally, no matter how well-intentioned, any form of sweeping government-sanctioned monitoring system poses important questions: how are our confidential data used? Who does have access to this? What is our data open to breaches and hacks? How will private corporations be using it in the future? And, obviously, is there a way to minimize the possibility of violations of privacy? There are critical issues that will most definitely resurface - even though we are too stressed now to worry about them - once in the post-coronavirus period hysteria ebbs and calm have been restored. Mobile location data provides advanced tracking capabilities for governments to help implement quarantine enforcement. Digital thermometers are being combined with face recognition technologies related to biometric repositories to better identify the identification of individuals with a fever. Open access tools, such as Nextstrain, use Gisaid, a forum for genomic data exchange, to help researchers identify and analyze the virus.

In an unprecedented situation several policymakers are able to neglect the ramifications of privacy in an attempt to save lives. The confidential data being gathered, though, is not limited to policy and public health organisations. Throughout the United States, the government is partnering publicly with Verily, a sister corporation owned by Google, to provide electronic verification services allowing users to have a Google account. Surveillance security providers and smartphone device creators are now allowing access to confidential data. For example, users of the Corona 100 m app will see the date a coronavirus patient was infected, along with their race, ethnicity, age and where they went. Sensitive medical records related to a patient will and should be kept confidential in ordinary circumstances. Exposing them to private corporations is a matter of anxiety, particularly in the name of public health, as such documents have considerable economic importance. For example, they could provide valuable targeting data for health-care and pharmaceutical companies to advertising agencies. They could also help inform health insurers making decisions seeking to verify medical history when processing new policies and claims. Databases containing identities linked to mobile location data do bear a price tag, especially for consumer markets.

Data protection laws such as the European General Data Protection Regulation (GDPR) and the California Consumer Privacy Act (CCPA) would limit businesses attempting to retain personal data of any kind, and even use it for potential commercial advantage. However, businesses must adopt the newest advances in privacy-enhancing technologies (PET) to completely ensure regulatory enforcement and secure data-an enormously important market commodity. As demonstrated by the World Economic Forum, this emerging type of privacy technologies helps companies to exploit knowledge obtained from third-party private data without disclosing sensitive information which can not and should not be exchanged. Luckily sophisticated cryptographic methods based on PET are now in use by businesses. The global research community has rigorously sought and tested them, and business members are deeply engaged in attempts to standardize PETs such as ZKProof to promote wider acceptance. If correctly applied, PET will inspire companies, rather than constrain them. It can help them to securely leverage third-party data and remain competitive, without jeopardizing user privacy or business confidentiality.

\section{\textbf{Conclusion}}
Preserving privacy as we develop and implement these technical solutions will be critical. It is more than important ever to consider privacy principles as we collectively move forward into this next phase of tracking, tracing and testing, and using similar technologies developed to address the pandemic. Meaningful consent must be obtained by being transparent about the reason for collecting data, what data is collected and how long it is kept. Only when people make the decision to participate should data be collected with consent and used in the manner explained. Clear and user-friendly communication aims to encourage cooperative engagement and will ensure that anyone engaging with the system makes informed decisions to engage in data collection. This should also ensure the user is aware of the purpose of gathering the data, the essence of the data to be collected, the length of data retention and the advantages of data collection.

\bibliographystyle{ieeetr}

\bibliography{malware,talukder}

\begin{thebibliography}{10}

\bibitem{fernandes2020economic}
N.~Fernandes, ``Economic effects of coronavirus outbreak (covid-19) on the
  world economy,'' {\em Available at SSRN 3557504}, 2020.

\bibitem{talukder2020tsc}
S.~Talukder and B.~Carbunar, ``A study of friend abuse perception in
  facebook,'' {\em Transactions on Social Computing}, vol.~1, no.~1, 2020.

\bibitem{ahani2020coronavirus}
A.~Ahani and M.~Nilashi, ``Coronavirus outbreak and its impacts on global
  economy: the role of social network sites,'' {\em Journal of Soft Computing
  and Decision Support Systems}, vol.~7, no.~2, pp.~19--22, 2020.

\bibitem{talukder2018abusniff}
S.~Talukder and B.~Carbunar, ``Abusniff: Automatic detection and defenses
  against abusive facebook friends,'' in {\em Twelfth International AAAI
  Conference on Web and Social Media}, 2018.

\bibitem{merchant2020social}
R.~M. Merchant and N.~Lurie, ``Social media and emergency preparedness in
  response to novel coronavirus,'' {\em Jama}, 2020.

\bibitem{talukder2019detection}
S.~K. Talukder, ``Detection and prevention of abuse in online social
  networks,'' {\em FIU Electronic Theses and Dissertations. 4026}, 2019.

\bibitem{sharfuddin2020.1760498}
S.~Sharfuddin, ``The world after covid-19,'' {\em The Round Table}, vol.~109,
  no.~3, pp.~247--257, 2020.

\bibitem{TSRICIEV14}
S.~K. Talukder, M.~I.~I. Sakib, and M.~M. Rahman, ``Model for e-government in
  bangladesh: A unique id based approach,'' in {\em 2014 International
  Conference on Informatics, Electronics Vision (ICIEV)}, pp.~1--6, May 2014.

\bibitem{okereafor2020tackling}
K.~Okereafor and O.~Adebola, ``Tackling the cybersecurity impacts of the
  coronavirus outbreak as a challenge to internet safety,'' {\em Journal
  Homepage: http://ijmr. net. in}, vol.~8, no.~2, 2020.

\bibitem{talukder2017usensewer}
S.~Talukder, M.~I.~I. Sakib, Z.~R. Talukder, U.~Das, A.~Saha, and N.~S.~N.
  Bayev, ``Usensewer: Ultrasonic sensor and gsm-arduino based automated
  sewerage management,'' in {\em 2017 International Conference on Current
  Trends in Computer, Electrical, Electronics and Communication (CTCEEC)},
  pp.~12--17, IEEE, 2017.

\bibitem{ahmad2020corona}
T.~Ahmad, ``Corona virus (covid-19) pandemic and work from home: Challenges of
  cybercrimes and cybersecurity,'' {\em Available at SSRN 3568830}, 2020.

\bibitem{talukdersurvey}
S.~Talukder and Z.~Talukder, ``A survey on malware detection and analysis
  tools,'' {\em International Journal of Network Security \& Its Applications},
  vol.~12, no.~2, 2020.

\bibitem{lajous2010mobile}
M.~Lajous, L.~Danon, R.~L{\'o}pez-Ridaura, C.~M. Astley, J.~C. Miller, S.~F.
  Dowell, J.~J. O’Hagan, E.~Goldstein, and M.~Lipsitch, ``Mobile messaging as
  surveillance tool during pandemic (h1n1) 2009, mexico,'' {\em Emerging
  infectious diseases}, vol.~16, no.~9, p.~1488, 2010.

\bibitem{talukder2020tools}
S.~Talukder, ``Tools and techniques for malware detection and analysis,'' {\em
  arXiv preprint arXiv:2002.06819}, 2020.

\bibitem{talukder2017attacks}
S.~Talukder, I.~I. Sakib, F.~Hossen, Z.~R. Talukder, and S.~Hossain, ``Attacks
  and defenses in mobile ip: Modeling with stochastic game petri net,'' in {\em
  2017 International Conference on Current Trends in Computer, Electrical,
  Electronics and Communication (CTCEEC)}, pp.~18--23, 2017.

\bibitem{TC2017}
S.~Talukder and B.~Carbunar, ``When friend becomes abuser: Evidence of friend
  abuse in facebook,'' in {\em {Proceedings of the 9th ACM Conference on Web
  Science}}, WebSci '17, (New York, NY, USA), ACM, June 2017.

\bibitem{talukder2019ccsc}
S.~Talukder, ``Abusniff: An automated social network abuse detection system,''
  {\em J. Comput. Sci. Coll.}, vol.~35, p.~209–210, Oct. 2019.

\bibitem{TSRICEEICT14}
S.~K. Talukder, M.~I.~I. Sakib, and M.~M. Rahman, ``Digital land management
  system: A new initiative for bangladesh,'' in {\em 2014 International
  Conference on Electrical Engineering and Information Communication
  Technology}, pp.~1--6, April 2014.

\bibitem{BBCapp}
L.~Kelion, ``Coronavirus: Nhs contact tracing app to target 80\% of smartphone
  users.'' BBC, \url{shorturl.at/qKT45}, 2020.

\bibitem{talukder2018mobile}
S.~Talukder, S.~Witherspoon, K.~Srivastava, and R.~Thompson, ``Mobile
  technology in healthcare environment: Security vulnerabilities and
  countermeasures,'' {\em arXiv:1807.11086}, 2018.

\bibitem{Unacast}
Unacast, ``Social distancing scoreboard.'' Unacast, \url{shorturl.at/gs068},
  2020.

\end{thebibliography}

\section*{AUTHORS}

\noindent
\begin{wrapfigure}{r}{0.2\linewidth}
\centering
\includegraphics[width=0.2\textwidth]{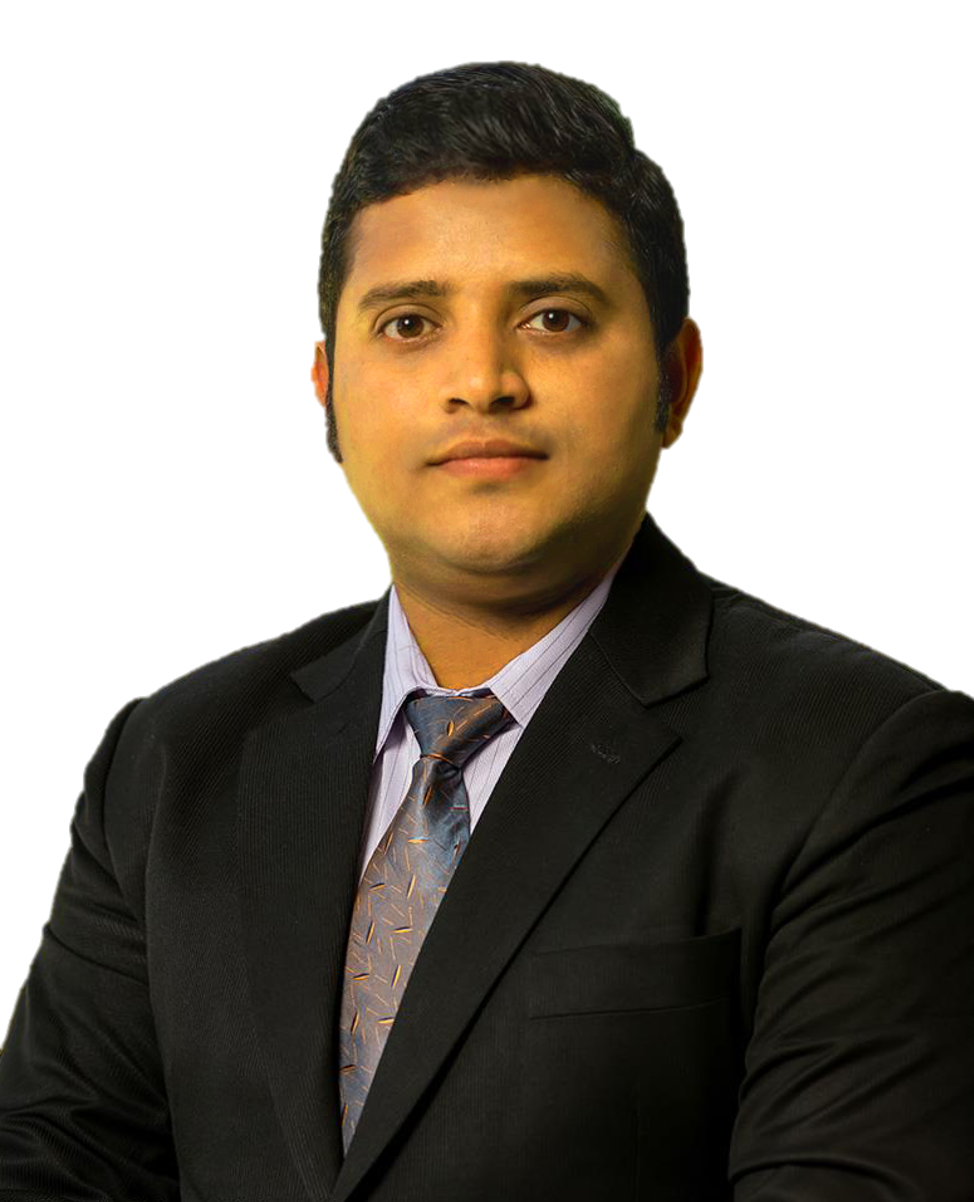}
\label{fig:myfig}
\end{wrapfigure}
\textbf{Sajedul Talukder, Ph.D.} is a tenure-track Assistant Professor of Computer Science at Edinboro University and the founder and director of \href{http://penslab.cs.edinboro.edu}{Privacy Enhanced Security Lab (PENSLab)}, where he and his group develop privacy enhanced security systems. Dr. Talukder's research interests include security and privacy with applications in online and geosocial networks, machine learning, wireless networks, distributed systems, and mobile applications. His research works have been published on top-tier social networking conferences and invited by Facebook in their headquarter. His work attracted a number of media attention including from NBC 6 and Sage Research Methods. In addition, he is also serving in the editorial board and program committee in several prestigious conferences and journals. Before joining EU, Dr. Talukder worked as a research mentor for Science without Borders, NSF-RET and NSF-REU at FIU.

\noindent
\begin{wrapfigure}{r}{0.2\linewidth}
\centering
\includegraphics[width=0.2\textwidth]{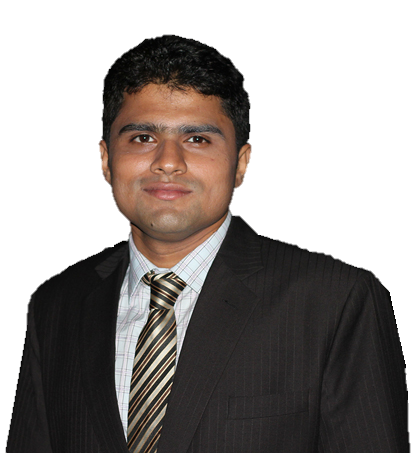}
\label{fig:myfig}
\end{wrapfigure}
\textbf{Md. Iftekharul Islam Sakib} is a Ph.D. student in Computer Science at the University of Illinois at Urbana-Champaign. Currently, he is working in Cyber Physical Computing group at UIUC and advised by Professor Tarek Abdelzaher. Before that, he served the Department of Computer Science and Engineering (CSE) at Bangladesh University of Engineering and Technology (BUET) as an Assistant Professor and currently in study leave for higher studies. He accomplished his M.Sc. \& B.Sc. from the same department in 2014 \& 2018 respectively.

\noindent
\begin{wrapfigure}{r}{0.2\linewidth}
\centering
\includegraphics[width=0.2\textwidth]{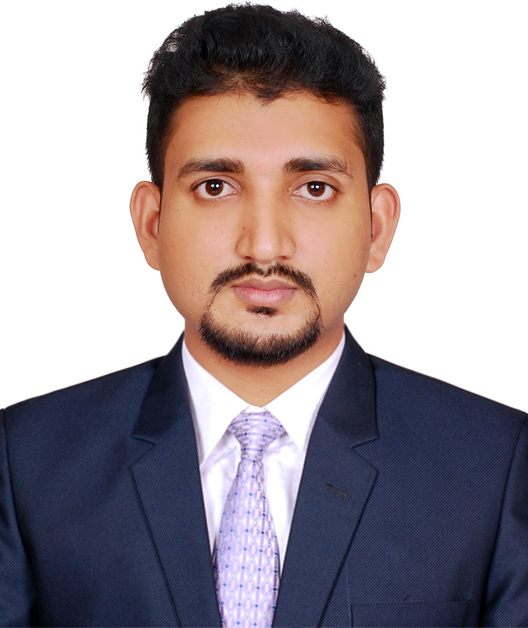}
\label{fig:myfig}
\end{wrapfigure}
\textbf{Zahidur Talukder} is a Ph.D. student in Computer Science at the University of Texas at Arlington. Currently, he is working in Rigorous Design Lab (RiDL) at UTA and advised by Professor Mohammad Atiqul Islam. His research interests are broadly in the areas of cyber-physical systems, computer architecture, and security. Currently, he is working on data center security, with a particular focus on mitigating the emerging threat of ``power attacks'' in multi-tenant ``colocation'' data centers.

\end{document}